\documentclass[aps,twocolumn,prc,amssymb,amsmath,floatfix,superscriptaddress,nofootinbib,float,romannum]{revtex4}

\usepackage[dvipsnames,usenames]{xcolor}
\usepackage{graphicx}
\usepackage{mathptmx}
\usepackage{bm}
\usepackage{setspace} 

\newcommand{\qq}{\mathbf{q}}
\newcommand{\pp}{\mathbf{p}}
\newcommand{\kk}{\mathbf{k}}

\newcommand{\be}{\begin{equation}}
\newcommand{\ee}{\end{equation}}
\newcommand{\bea}{\begin{eqnarray}}
\newcommand{\eea}{\end{eqnarray}}
\newcommand{\ba}{\begin{align}}
\newcommand{\ea}{\end{align}}

\newcommand{\rn}{{\rm n}}
\newcommand{\rp}{{\rm p}}
\newcommand{\ra}{{\rm a}}
\newcommand{\rb}{{\rm b}}
\newcommand{\rc}{{\rm c}}
\newcommand{\rd}{{\rm d}}
\newcommand{\rs}{{\rm s}}

\newcommand{\up}{\uparrow}
\newcommand{\down}{\downarrow}

\begin{document}

\title{Multicomponent Fermi systems at low densities}

\author{C. J. Pethick}
\email{pethick@nbi.dk}
\affiliation{The Niels Bohr International Academy, The Niels Bohr Institute, University of Copenhagen, Blegdamsvej 17, DK-2100 Copenhagen \O, Denmark}
\affiliation{NORDITA, KTH Royal Institute of Technology and Stockholm University, Hannes Alfv\'ens v\"ag 12, SE-106 91 Stockholm, Sweden}

\author{A. Schwenk}
\email{schwenk@physik.tu-darmstadt.de}
\affiliation{Technische Universit\"at Darmstadt, Department of Physics, D-64289 Darmstadt, Germany} 
\affiliation{ExtreMe Matter Institute EMMI, GSI Helmholtzzentrum f\"ur Schwerionenforschung GmbH, D-64291 Darmstadt, Germany}
\affiliation{Max-Planck-Institut f\"ur Kernphysik, Saupfercheckweg 1, D-69117 Heidelberg, Germany}

\begin{abstract}
We calculate, to second order in the scattering length between two fermions, the Landau quasiparticle interaction for a low-density mixture of two fermion species with unequal densities at temperature zero. From the Landau parameters we evaluate the energy density and find agreement with the result of Kanno, Prog. Theor. Phys. {\bf 44}, 813 (1970). The calculations are then extended to the case of two fermion components with different total densities, each with two spin components, a situation of interest in nuclear physics and astrophysics, where the species are neutrons and protons. An interesting finding is that, for low proton concentrations, $x \ll 1$, the leading term in the energy density, beyond the $x^{5/3}$ contribution from the kinetic energy and the $x^2$ one due to the two-body interaction in the mean-field approximation, varies as $x^{7/3} \ln x$. This is to be contrasted with the higher powers of $x$ implicit in many phenomenological energy-density functionals employed in nuclear physics, such as those of the Skyrme type.
\end{abstract}

\maketitle

\section{Introduction}

This work is motivated by the desire to improve the description of neutron-rich matter such as that found in neutron stars. There the proton fraction $x=n_\rp/(n_\rn + n_\rp)$, where $n_\rn$ is the neutron density and $n_\rp$ the proton density, is typically of order 5\% at densities around nuclear saturation density. To this end, we wish to elucidate the behavior of the energy density of bulk matter consisting of neutrons and protons at low total baryon density, $n=n_\rn + n_\rp$.
As we shall show, for low proton concentration, the analytic structure of the energy density differs from that given by many widely used phenomenological energy functionals, such as those of the Skyrme type~\cite{StoneReinhard}.

Study of degenerate Fermi systems with two components with different densities has a long history dating back to Stoner's model for ferromagnetism~\cite{Stoner}. In the 1960s, following the development of Landau's theory of Fermi liquids~\cite{Landau1,Landau2,Landau3,BaymPethick,Migdal,FrimanHebelerSchwenk}, there were numerous studies of ferromagnetism based on such a model, both in the context of laboratory materials as well as in astrophysics, where it had been suggested that the high magnetic fields in pulsars, which had been discovered at that time, could be a consequence of a ferromagnetic state of neutrons. In the astrophysical setting, Kanno~\cite{Kanno} extended to the case of two components with unequal densities the calculations of the ground-state energy, which had been performed for a low-density gas with two components with equal densities~\cite{HuangYang, DeDominicisMartin,AbrikosovKhalatnikov}.\footnote{A textbook exposition of the work of Ref.~\cite{AbrikosovKhalatnikov} may be found in Ref.~\cite{LifshitzPitaevskii}.} While for two components with equal densities, the calculations have been carried to high orders in an effective field theory (EFT) expansion~\cite{HammerFurnstahl,WDS2020,WDS2021}, analytical results for two components of different densities so far are limited to the terms up to second order in the scattering length, and in this paper we shall confine ourselves to this case. 

More recently, binary mixtures of fermions have received renewed attention as a result of the possibility of making experiments on low-density mixtures of fermionic atoms. Much of this has been numerical, because of the difficulty of performing analytical calculations when the two fermions have different masses~\cite{FratiniPilati,ChankowskiWojtkiewicz1,ChankowskiWojtkiewicz2}. Recently, Pera et al. have presented a direct evaluation of the second-order contribution to the energy of an SU(N) Fermi gas~\cite{Pera}.

In this paper we begin by revisiting the calculation of the energy density of a Fermi gas with two components with different Fermi momenta. We shall consider the case of equal masses. Kanno reported his results only in a short communication \cite{Kanno}, and we shall provide a detailed derivation of the expression for the energy density. The approach we adopt differs from Kanno's in that we first calculate the Landau parameters, which have not previously been evaluated, and from them calculate the energy density. This is the procedure used by Abrikosov and Khalatnikov~\cite{AbrikosovKhalatnikov} for the case of equal Fermi momenta and it simplifies integrations compared with the direct evaluation of integrals performed by Kanno.

In Sec.~II we calculate the sound speed of a binary mixture of fermions in terms of Landau parameters and from that we shall derive an expression for the contribution to the energy density to second order in the scattering length. Section III is devoted to the calculation of the quasiparticle interaction, Landau parameters, and energy density for a binary fermion mixture, and in Sec. IV the results are generalized to the case of neutron--proton mixtures. Section V contains a discussion of the results and their implications. The calculational details are given in Appendix A.

\section{Sound velocity and Landau parameters}

In this section we relate the sound velocity to the Landau parameters, and from that calculate the contribution to the energy to second order in the scattering length. 
We consider two components, which we shall denote by ``a'' and ``b'', with densities $n_\ra$ and $n_\rb$ and the same mass $m$. It is convenient to consider the energy density as a function of the total density $n=n_\ra+ n_\rb$, and the concentration $x_\ra=n_\ra/n$. The concentrations satisfy the condition $x_\ra+x_\rb=1$. In the nonrelativistic limit, the quantity $\partial^2 E/\partial n^2|_{x_\ra}$, $E$ being the total energy density, is related to the first sound velocity, $s$, in the medium by the relation
\be
\frac{\partial^2 E}{\partial n^2} =\frac{\partial \mu}{\partial n}=\frac{ms^2}{n} \,,
\ee 
where $\mu$, the total chemical potential is given in terms of the chemical potentials $\mu_\ra$ and $\mu_\rb$ by $\mu=x_\ra\mu_\ra +x_\rb\mu_\rb$.
The energy density can then be obtained by two integrations with respect to the density. The advantages of this way of proceeding are twofold: first, the sound velocity may be directly expressed in terms of the Landau quasiparticle interaction, and therefore it is possible to identify the regions of phase space that are responsible for particular terms in the expression for the energy, and second, the integrations with respect to the density are straightforward. 

In this work, we assume the system is in a normal ground state, i.e., we neglect superfluid pairing correlations and we assume there are no bound states. Our work is thus applicable where pairing gaps are small compared with the Fermi energy, so that pairing contributions to the energy density will be small (see, e.g., Ref.~\cite{Zwerger}).  

We consider variations of the densities of both components, $\delta n_\ra$ and $\delta n_\rb$, so that the ratio of a-particles to b-particles remains the same. Therefore, we have
\be
\frac{\delta n_\ra}{n_\ra} =\frac{\delta n_\rb}{n_\rb}=\frac{\delta n}{n} \,.
\label{deltan}
\ee
The energy change to second order in the changes $\delta n_\ra$ and $\delta n_\rb$ in the densities, and no angular distortions of the Fermi surfaces is given by
\begin{align}
&\delta^2E= \frac{(\delta n_\ra)^2}{2N_\ra(0)} + \frac{(\delta n_\rb)^2}{2N_\rb(0)} \nonumber \\
&+\frac12 f_0^{\rm aa} (\delta n_\ra)^2 + \frac12 f_0^{\rm bb} (\delta n_\rb)^2 +f_0^{\rm ab} \delta n_\ra \delta n_\rb \,,
\label{d2E}
\end{align}
where
\be
N_\ra(0)= \frac{m_\ra^* p_\ra}{2\pi^2}
\quad {\rm and} \quad N_\rb(0)= \frac{m_\rb^* p_\rb}{2\pi^2}
\ee
are the densities of quasiparticle states at the Fermi surface, $p_\ra$ and $p_\rb$ being the respective Fermi momenta and $m_\ra$ and $m_\rb$ the effective masses. The densities are given in terms of the Fermi momenta by the expression
\be
n_\alpha=\frac{p_\alpha^3}{6 \pi^2} \,,
\ee
where we have suppressed factors of $\hbar$, as we shall throughout this work.
The Landau quasiparticle interaction is defined by
\be
f^{\alpha\beta}(\pp, \pp')=\frac{\delta^2 E}{\delta n_{\pp\alpha} \delta n_{\pp' \beta}} \,, 
\ee
where $\alpha$ and $\beta$ are species labels and $n_{\pp\alpha}$ is the quasiparticle occupation number for momentum state $\pp$ and species $\alpha$. For $\pp$ and $\pp'$ on the respective Fermi surfaces, one may expand $f$ is terms of Legendre polynomials, $P_\ell(\zeta)$, where $\zeta=\hat\pp \cdot\hat \pp'$:
\be
f^{\alpha\beta}(\pp, \pp')=\sum_\ell f^{\alpha\beta}_\ell P_\ell(\zeta) \,.
\label{LandauParam}
\ee
As shown by Sj\"oberg, for the case of equal masses, $m_\ra=m_\rb=m$, one can write the effective masses in the form \cite{Sjoeberg} 
\be
\frac{m_\ra^*}{m}=1+\frac{F_1^\ra}{3} \,,
\ee
where
\be
F_1^\ra=N_\ra(0) \left(f_1^{\ra\ra} + f_1^{\ra\rb}\frac{p_\rb^2}{p_\ra^2}\right),
\ee
and similar expressions with the roles of a- and b-particles interchanged.
Thus it follows that 
\be
\frac{1}{N_\ra(0)}=\frac{1}{N_\ra^0(0)}-\frac13 \left( f_1^{\ra\ra} + f_1^{\ra\rb}\frac{p_\rb^2}{p_\ra^2} \right),
\label{dos}
\ee
where{\footnote{The factor of 2 difference between Sj\"oberg's expressions for densities of states and ours is due to the fact that we work in terms of the density of states for a single internal (spin) state, whereas his results are for nucleons, each with two spin states.}
\be
N_\ra^0(0)=\frac{m p_\ra}{2\pi^2}
\ee
is the free-particle density of states at the Fermi surface. The effective mass is thus given by
\be
\frac{m}{m_\ra^*}=1-\frac{N_\ra^0(0)}3 \left( f_1^{\ra\ra} + f_1^{\ra\rb}\frac{p_\rb^2}{p_\ra^2} \right).
\ee
On combining Eqs.~(\ref{deltan}), (\ref{d2E}) and (\ref{dos}), one finds
\begin{align}
\frac{\partial \mu}{\partial n}= \frac{ms^2}{n} &=\left[\frac{1}{N_\ra^0(0)}+f_0^{\rm a a}-\frac13 f_1^{\rm a a}\right]x_\ra^2 \nonumber \\ &+ \left[\frac{1}{N_\rb^0(0)}+f_0^{\rm b b} -\frac13 f_1^{\rm b b}\right]x_\rb^2 \nonumber \\
&+ \left[2f_0^{\rm a b}- \frac{f_1^{\rm a b}}{3} \left\{\left(\frac{x_\ra}{x_\rb}\right)^{1/3} + \left(\frac{x_\rb}{x_\ra}\right)^{1/3} \right\} \right]x_\ra x_\rb \,.
\end{align}

In this work, we shall be particularly interested in terms of order $a^2$, where $a$ is the free-space scattering length and we shall indicate these by a superscript ``$(2)$''. The integrals in the various expressions all converge, so on dimensional grounds it follows that at given concentrations of a- and b-particles, the second-order contribution to $ms^2/n$ varies as $n^{1/3}$  times a function of the concentrations, as we shall confirm later by detailed calculations.
Therefore, after integrating twice with respect to $n$, one sees that the corresponding contribution to the energy density is $(9/28) n^2$ times the contribution to $\partial \mu/\partial n$:
\begin{align}
E^{(2)} & =\frac{9}{28}\left\{ \left[ f_0^{ (2)\ra \ra}-\frac13 f_1^{ (2)\ra \ra}\right]n_\ra^2 +\left[ f_0^{ (2)\rb \rb}-\frac13 f_1^{ (2)\rb \rb}\right]n_\rb^2 \right.\nonumber\\& + \left.\left[2f_0^{ (2)\ra \rb}- \frac{f_1^{ (2)\ra \rb}}{3} \left(\frac{p_\ra^2+p_\rb^2}{p_\ra p_\rb} \right) \right]n_\ra n_\rb\right\}\\
&=\frac{1}{112\pi^4}\left\{ \left[ f_0^{ (2)\ra \ra}-\frac13 f_1^{ (2)\ra \ra}\right]p_\ra^6 +\left[ f_0^{ (2)\rb \rb}-\frac13 f_1^{ (2)\rb \rb}\right]p_\rb^6 \right.\nonumber\\& + \left.\left[2f_0^{ (2)\ra \rb}- \frac{f_1^{ (2)\ra \rb}}{3} \left(\frac{p_\ra^2+p_\rb^2}{p_\ra p_\rb} \right) \right]p_\ra^3 p_\rb^3\right\}.
\label{E(2)}
\end{align}

\section{Quasiparticle interaction}
\label{qpinteraction}

In this section we calculate the quasiparticle interaction and Landau parameters to second order in the scattering length for the two species. We shall consider here the case when the states a and b correspond to the two spin-1/2 substates of the same particle. In that case, the Pauli principle implies that the scattering length for like particles vanishes (see, e.g., Ref.~\cite{LifshitzPitaevskii}). The contribution to the energy density of first order in the scattering length is
\be
E^{(1)}=U_0\sum_{1,2}n_{1\ra}n_{2\rb} \,.
\ee
The second-order contribution is given by\footnote{This equation is the starting point of Kanno's calculation \cite{Kanno} for a hard-sphere Fermi gas. However, the results are more general and may also be derived in the language of effective field theory~\cite{HammerFurnstahl,ChankowskiWojtkiewicz1}. } 
\be
E^{(2)} =- \sum_{1,2,3,4} U_0^2 \frac{n_{1\ra}n_{2\rb}(n_{3\ra}+n_{4\rb})}{\epsilon_1 +\epsilon_2 -\epsilon_3 -\epsilon_4}\delta (\pp_1+\pp_2-\pp_3-\pp_4) \,,
\label{Energyab}
\ee
where $\epsilon_i=p_i^2/2m$, $U_0=4 \pi a/m$, $a$ being the a--b scattering length, and in the sum over momentum states in unit volume we use $1$ as a shorthand for $\pp_1$, and so on. The quantity $n_{i,\alpha}$ is the distribution function for particles of species $\alpha$ and momentum $\pp_i$, which in the ground state is unity for $p_i<p_\alpha$, $p_\alpha$ being the Fermi momentum for species $\alpha$, and zero otherwise.
The energy denominator in Eq.\ (\ref{Energyab}) can vanish and the integrals must be interpreted as principal value ones, as is explained in detail in Ref.~\cite{ChankowskiWojtkiewicz1}.

The first-order contributions to the Landau parameters are given by
\be
f^{(1)\rm ab}(\pp, \pp')=U_0,\,\,\,({\rm a}\neq {\rm b})
\label{f^(1)}
\ee
and they vanish if $\rm a={\rm b}$.
These results are independent of the relative populations of the two components and independent of the momenta.

The second-order contribution to the Landau $f$ function is given for two a-particles on their Fermi surface by
\begin{align}
f^{(2)\rm aa}(\pp, \pp')
&=-U_0^2\sum_{1,2}\frac{2 n_{1\rb}\delta (\pp -\pp'-\pp_1+\pp_2)}{\epsilon_1 -\epsilon_2} \\
&=-2m U_0^2\sum_1\frac{2 n_{1\rb}}{p_1^2 -(\pp_1-\qq)^2} \label{Lindhardintegral}\\
&=U_0^2 \chi^{\rm bb}(q) \,,
\label{faa}
\end{align}
where
\be
\qq=\pp-\pp'
\label{q}
\ee
is the relative momentum of the two particles, 
and
\be
\chi^{\alpha\alpha}(q)= \frac{m p_\alpha}{2\pi^2}\left(\frac12+ \frac{1-y_\alpha^2}{4y_\alpha} \ln\left|\frac{1+y_\alpha}{1-y_\alpha} \right|\right)
\label{staticLindhard}
\ee
is the static Lindhard function with $y_\alpha=q/(2p_\alpha)$. While considering neutrons and protons, we have taken their masses to be equal but, in the case of unequal masses, $m$ must be replaced by $m_\alpha$. Physically Eq.\ (\ref{faa}) represents the interaction between two a-particles induced by exchange of a b-particle--b-hole pair. While the second-order contribution to the energy might be expected to be negative, it gives a positive contribution to the quasiparticle interaction because it corresponds to an exchange term.
For two b-particles, the result is
\be
f^{(2)\rm bb}(\pp, \pp')=U_0^2 \chi^{\rm aa}(q)\,.
\label{fbb}
\ee

For interaction of an a-particle with a b-particle, both on their respective Fermi surfaces, the result is
\begin{align}
f^{(2)\ra\rb}({\bf{p, p'}})&= -2mU_0^2\sum_{1}\frac{(n_{1\ra}+n_{1\rb})}{p_\ra^2 +p_\rb^2 -p_1^2 -(\kk-\pp_1)^2}\nonumber\\
&-2mU_0^2\sum_1\frac{n_{1\ra} }{p_\ra^2 -p_\rb^2 + p_1^2 -(\pp_1-\qq)^2}\nonumber\\
&-2mU_0^2\sum_1\frac{n_{1\rb}}{-(p_\ra^2 -p_\rb^2) + p_1^2 -(\pp_1-\qq)^2} \,,
\label{fabbasic}
\end{align} 
where
\be
\kk=\pp+\pp'
\label{k}
\ee
is the total momentum. For $\pp$ and $\pp'$ lying on the respective Fermi surfaces, it follows that
\be
q^2+k^2=2(p_\alpha^2+p_\beta^2)\,,
\label{qk}
\ee
and therefore the variables $\qq$ and $\kk$ are not independent, and we shall write terms as functions of $\qq$ or $\kk$ according to which choice is the more convenient.
The first term in Eq.~(\ref{fabbasic}) represents a reduction of the second-order contribution to the energy due to blocking of intermediate two-particle states by the filled Fermi seas and the second term is due to exchange of an a--b particle--hole pair. Details of the calculations are given in Appendix A and the result is
\begin{align}
&f^{(2)\ra\rb}(q)=\frac{mU_0^2}{2\pi^2}
\left[ \frac{p_\ra+p_\rb}4 \left( 3+\frac{(p_\ra-p_\rb)^2}{q^2} \right)\right. \nonumber \\
&- \left(\frac{(p_\ra^2-p_\rb^2)^2}{8q^3}-\frac{( p_\ra^2+p_\rb^2)}{4q} +\frac{3 q}{8} \right) \ln \left| \frac{q+p_\ra+p_\rb}{q-p_\ra-p_\rb} \right| \nonumber \\
&+ \left.\frac{\left(p_\ra^2 - p_\rb^2 \right)}{4k}
\ln \left| \frac{k+p_\ra -p_\rb}{k-p_\ra +p_\rb} \right| \right].
\label{fabq}
\end{align}

\subsection{Landau parameters}

From the results above one can calculate the angular moments. From Eq.~(\ref{f^(1)}), one sees that the first-order contribution to the Landau parameters has only an $l=0$ component given by
\be
f^{(1)\ra\rb}_0=U_0 (1-\delta_{\ra\rb}) \,.
\ee
The second-order contributions are given by Eqs.~(\ref{faa}), (\ref{fbb}), and (\ref{fabq}) and details of the calculations are given in Appendix~\ref{details}. One finds
\begin{align}
f^{(2)\ra\ra}_0&=\frac{mU_0^2}{2\pi^2}\left[ \frac{p_\rb}3 
+\left( -\frac{p_\ra}{6}+\frac{p_\rb^2}{2p_\ra} \right) \ln\left| \frac{p_\ra + p_\rb}{p_\ra - p_\rb} \right| \right.\nonumber\\ 
&\left. +\frac{p_\rb^3}{3p_\ra^2} \ln\left|\frac{p_\ra^2-p_\rb^2}{p_\rb^2} \right| \right],
\label{faa0final}
\end{align}
\begin{align} 
\frac13 f^{(2)\ra\ra}_1 &= \frac{mU_0^2}{2\pi^2}\left[ -\frac{p_\rb}{15} -\frac{2}{15}\frac{p_\rb^3}{p_\ra^2} \right. +\left(\frac{p_\ra}{30} +\frac{p_\rb^2}{6p_\ra}\right) \ln\left| \frac{p_\ra + p_\rb}{p_\ra-p_\rb} \right| \nonumber \\ 
& \left. + \left( \frac{ p_\rb^3}{3p_\ra^2}-\frac{2}{15}\frac{p_\rb^5}{p_\ra^4} \right) \ln\left| \frac{p_\ra^2-p_\rb^2}{p_\rb^2} \right| \right],
\label{faa1final}
\end{align}
and the combination that occurs in the energy density is
\begin{align} 
&f^{(2)\ra\ra}_0-\frac13 f^{(2)\ra\ra}_1= \frac{mU_0^2}{2\pi^2} \left[ \frac25 p_\rb +\frac{2}{15}\frac{p_\rb^3}{p_\ra^2}\right. \nonumber\\ 
&\left. -\left( \frac{p_\ra}{5} -\frac{p_\rb^2}{3 p_\ra}\right) \ln\left| \frac{p_\ra +p_\rb}{p_\ra - p_\rb} \right| +\frac{2}{15}\frac{p_\rb^5}{p_\ra^4} \ln\left| \frac{p_\ra^2-p_\rb^2}{p_\rb^2} \right| \right].
\end{align}
The corresponding results for $f^{(2)\rb\rb}_0$ and
$f^{(2)\rb\rb}_1$ may be obtained from these by exchanging a and b.
For the interactions between quasiparticles in different spin states one finds
\begin{align} 
&f^{(2)\ra\rb}_0=\frac{mU_0^2}{2\pi^2}(p_\ra+p_\rb)\times \nonumber \\
&\left[ \frac12 
+\frac{(p_\ra - p_\rb)}{4 p_\ra p_\rb}\left(p_\ra\ln\left|\frac{p_\ra^2-p_\rb^2}{p_\ra^2}\right| - p_\rb\ln\left|\frac{p_\ra^2-p_\rb^2}{p_\rb^2}\right| \right)\right]
\label{fab0final}
\end{align}
and
\begin{align} 
&\frac13 f^{(2)\ra\rb}_1= \frac{mU_0^2}{2\pi^2}
\left[ \frac{2}{15}(p_\ra + p_\rb)- \frac{p_\ra^3 +p_\rb^3}{10p_\ra p_\rb} \right. \nonumber \\
&+\left(\frac{(p_\ra +p_\rb)}{12}+\frac{1}{20}\frac{( p_\ra^5 + p_\rb^5 )}{p_\ra^2 p_\rb^2} \right)\ln\left|\frac{p_\ra+p_\rb}{p_\ra-p_\rb}\right|\nonumber \\ 
& \left.+ \frac{2}{15} \frac{1}{p_\ra^2p_\rb^2}\left(p_\ra^5\ln\left|\frac{p_\ra^2-p_\rb^2}{p_\ra^2}\right| +p_\rb^5\ln\left|\frac{p_\ra^2-p_\rb^2}{p_\rb^2}\right|\right) \right].
\label{fab1final}
\end{align}
For the unpolarized spin-1/2 system ($p_\ra = p_\rb$), our result agrees with the effective mass to second order in the dilute gas expansion~\cite{Galitskii,Platter2003}:\footnote{Note the sign typo in Eq.~(32) of Ref.~\cite{Platter2003}.}
\be
\frac{m^*}{m} = 1 + \frac{8}{15 \pi^2} (7 \ln2 -1) (k_{\rm F} a)^2 \,.
\ee

\subsection{Energy density}

On inserting the expressions for the Landau parameters into Eq.~(\ref{E(2)}), one finds
\be
E^{(2)}=U_0^2\Phi(p_\ra,p_\rb) \,,
\label{E2twocomp}
\ee
where
\begin{align}
&\Phi(p_\ra,p_\rb)=\frac{m}{3360\pi^6}\left[ 22p_\ra^3 p_\rb^3(p_\ra +p_\rb) \phantom{\frac12} \right. \nonumber \\ 
& +\frac12 (p_\ra-p_\rb)^2 p_\ra p_\rb(p_\ra+p_\rb) (15p_\ra^2+15 p_\rb^2+11p_\ra p_\rb)\nonumber \\ 
&-\frac74(p_\ra-p_\rb)^4(p_\ra+p_\rb)\left(p_\ra^2 + p_\rb^2+3 p_\ra p_\rb\right) \ln \left|\frac{p_\ra+p_\rb}{p_\ra -p_\rb}\right| \nonumber \\ 
& \left.-4p_\ra^7 \ln \frac{p_\ra +p_\rb}{p_\ra}-4p_\rb^7 \ln \frac{p_\ra +p_\rb}{p_\rb}\right] .
\label{Kanno}
\end{align}
This result agrees with that reported by Kanno \cite{Kanno}, which was confirmed in Ref.~\cite{Pera}. For $p_\ra=p_\rb=p_{\rm F}$ one recovers the well-known result \cite{LifshitzPitaevskii}
\be
E^{(2)}=\frac{mU_0^2p_{\rm F}^7}{840\pi^6} ( 11-2\ln 2 ) \,,
\label{E2eFS}
\ee
and for $p_\rb/p_\ra \rightarrow 0$ or $p_\ra/p_\rb \rightarrow 0$ the interaction tends to zero, a consequence of the Pauli principle.

In the nuclear case, one often uses a quadratic approximation to interpolate between the symmetric ($p_\ra=p_\rb$ or $x=1/2$) and the one-component system ($p_\rb=0$, $x=0$ or $p_\ra=0$, $x=1$). In order to compare our full second-order result, Eq.~(\ref{E2twocomp}), with such a quadratic interpolation, we study the ratio $E^{(2)}(n,x)/E^{(2)}(n,x=1/2) = \Phi(p_\ra,p_\rb)/\Phi(p_\ra=p_\rb)$ of the full result to the symmetric case, Eq.~(\ref{E2eFS}), as a function of $x=n_\rb/n$, so that the ratio is independent of total density $n=n_\ra+n_\rb$. This is plotted in Fig.~\ref{fig:quadratic}, where the curvature of the quadratic approximation has been taken to be equal to that of the full expression at $x=0.5$. The figure shows the shortcomings of the quadratic approximation, which underestimates the difference in between the energy densities for $x=0$ and $x=0.5$ by $\sim 13\%$. In particular, the quadratic approximation does not recover the result that $E^{(2)}$ vanishes for $x=0$ or 1 due to the Pauli principle for $S$-wave contact interactions. This could be remedied by adjusting the quadratic approximation to fit the extremes $x=0$ (or $x=1$) and $x=1/2$, but then the approximation would perform worse for other values of $x$.

\begin{figure}[t!]
\centering
\includegraphics[width=\columnwidth,clip=]{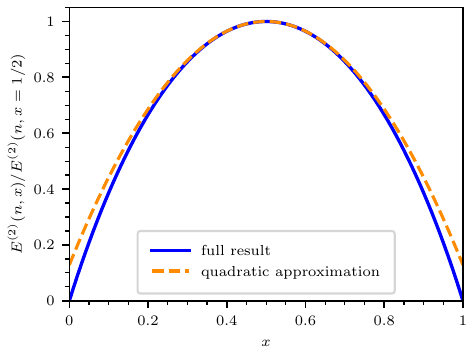}
\caption{Plot of the ratio of $E^{(2)}(n,x)$, Eq.~(\ref{E2twocomp}), to its value $E^{(2)}(n,1/2)$, Eq.~(\ref{E2eFS}), for matter with equal numbers of the two constituents, as well as the corresponding result for the quadratic approximation, as a function of $x$.}
\label{fig:quadratic}
\end{figure}

\section{Mixtures of neutrons and protons}

We now turn to mixtures of neutrons and protons which are not magnetically polarized. At the very lowest densities, the system is strictly speaking not a normal Fermi liquid because of a bound state, the deuteron, in the neutron--proton spin triplet channel, as well as other many-body bound states. Such matter consists of a mixture of neutrons, protons and bound states. However, at densities greater than the average density in the deuteron, the deuteron is dissolved, and similarly for other composite nuclei. It is then a reasonable first approximation to treat spatially uniform matter as a low-density normal Fermi liquid with effective two-body interactions which are not simply related to the free-space scattering lengths. From the expression for the energy density of a two-component system, it is straightforward to calculate the energy density of neutrons and protons with two spin states. There are 4 spin and isospin states, which we denote by a, b, c, and d. 

There are four independent effective low-energy interactions. 
Two of these are for like particles, and they are
\be
U_{\rn \uparrow, \rn \downarrow}^{\rn \uparrow, \rn \downarrow}=U_{\rn \downarrow, \rn \uparrow}^{\rn \downarrow, \rn \uparrow}=U_{\rn\rn}=\frac{4 \pi 
a_{\rn\rn}^{\uparrow\downarrow}}{m} 
\ee
for neutrons, where $a_{\rn\rn}^{\uparrow\downarrow}$ is the corresponding scattering length, and
\be
U_{\rp \uparrow, \rp \downarrow}^{\rp \uparrow, \rp \downarrow}=U_{\rp \downarrow, \rp \uparrow}^{\rp \downarrow, \rp \uparrow}=U_{\rp\rp}=\frac{4 \pi  a_{\rp\rp}^{\uparrow\downarrow}}{m} 
\ee
for protons, $a_{\rp\rp}^{\uparrow\downarrow}$ being the corresponding scattering length. For like spins the effective interactions vanish, as in the two-component case considered in Sec.~\ref{qpinteraction}.

The two other nonzero effective interactions are for the case when a and b refer to a neutron and a proton, for which there are four possible spin--isospin states. If a and b refer to the same spin state, spin conservation implies that the states c and d must have the same spin component, and the effective interaction corresponds to that for a neutron and a proton in a spin-triplet state, 
\be U_{\rn \uparrow, \rp \uparrow}^{\rn \uparrow, \rp \uparrow}=U_{\rn \downarrow, \rp \downarrow}^{\rn \downarrow, \rp \downarrow}
=U_{\rn\rp}^{S=1}\,.
\ee

When a and b correspond to different spin states, the states c and d can be either the same as a and b or the states with the spins interchanged. The effective interaction for the first possibility is 
\be U_{\rn \uparrow, \rp \downarrow}^{\rn \uparrow, \rp \downarrow}=U_{\rn \downarrow, \rp \uparrow}^{\rn \downarrow, \rp \uparrow}
=\frac12 (U_{\rn\rp}^{S=1}+U_{\rn\rp}^{S=0})\equiv U_{\rn\rp}^{\uparrow\downarrow}\,,
\ee
and that for the second one is 
\be U_{\rn \uparrow, \rp \downarrow}^{\rn \downarrow, \rp \uparrow}=U_{\rn \downarrow, \rp \uparrow}^{\rn \uparrow, \rp \downarrow}
=\frac12 (U_{\rn\rp}^{S=1}-U_{\rn\rp}^{S=0}) \equiv \tilde U_{\rn\rp}^{\uparrow\downarrow}\,.
\ee
In terms of the scattering lengths for neutron--proton scattering, $a^S_{\rn\rp}$,
\be
U_{\rn\rp}^S =\frac{4 \pi  a_{\rn\rp}^S}{m}\,. 
\ee
A summary of these results is given in Table I.

\begin{table}[t!]
\begin{center}
\begin{tabular}{|c|c|c|}
\hline
{\rm a b} & {\rm c d}& $U_{\rm a,b}^{\rm c,d}$\\
\hline
\rn$\uparrow$ \rn$\uparrow$ & \rn$\uparrow$ \rn$\uparrow$ & 0\\
\rn$\uparrow$ \rn$\downarrow$ & \rn$\uparrow$ \rn$\downarrow$ & $U_{\rm nn}^{\uparrow\downarrow}$\\
\hline
\rp$\uparrow$ \rp$\uparrow$ & \rp$\uparrow$ \rp$\uparrow$ & 0\\
\rp$\uparrow$ \rp$\downarrow$ & \rp$\uparrow$ \rp$\downarrow$ & $U_{\rm pp}^{\uparrow\downarrow}$\\
\hline
\rn$\uparrow$ \rp$\uparrow$ & \rn$\uparrow$ \rp$\uparrow$ & $U_{\rm np}^{\up\up}$\\
\rn$\uparrow$ \rp$\downarrow$ & \rn$\uparrow$ \rp$\downarrow$ & $U_{\rn\rp}^{\up\down}$\\
\rn$\uparrow$ \rp$\downarrow$ & \rn$\downarrow$ \rp$\uparrow$ & $\tilde U_{\rn\rp}^{\up\down}$\\
\hline
\end{tabular}
\caption{Matrix elements for nucleon--nucleon scattering. If the spin directions of all nucleons in a scattering processes are reversed, the matrix element is unchanged.}
\label{table1}
\end{center}
\end{table}

The first-order contribution to the energy density for a system that is not magnetically polarized is given by
\be
E^{(1)}=\frac14 U_{\rn\rn}n_\rn^2+\frac14 U_{\rp\rp}n_\rp^2+\frac18(U_{\rn\rp}^{S=0}+3U_{\rn\rp}^{S=1})n_\rn n_\rp\,.
\ee

The second-order contribution to the energy density obtained by generalizing Eq.~(\ref{Energyab}) to four internal states is
\begin{align}
&E^{(2)} =\nonumber \\
& \sum_{\{\ra,\rb, \rc,\rd\}}\sum_{1,2,3,4} (U_{\ra,\rb}^{\rc,\rd})^2 \frac{n_{1\ra}n_{2\rb}(n_{3\rc}+n_{4\rd})}{\epsilon_1 +\epsilon_2 -\epsilon_3 -\epsilon_4}\delta (\pp_1+\pp_2-\pp_3-\pp_4)\,,
\label{Energyabcd}
\end{align}
where the sum over species is to be taken over independent choices of the pairs of states a,b and c,d. Interactions between nucleons in the same spin and isospin state vanish by the Pauli principle, so there is a total of 6 ways of choosing two nucleons in different spin--isospin states. 
 
If all the states a-d refer to neutrons, the contribution to the energy density is given by that for a two-component Fermi gas with equal Fermi surfaces for the two spin components, Eqs.~(\ref{E2twocomp}) and (\ref{E2eFS}) with $U_0$ replaced by $U_{\rn\rn}$, the effective interaction between neutrons in different spin states. Corresponding results apply when all states refer to protons. 

On squaring the matrix elements for neutron--proton interactions in Eq.~(\ref{Energyabcd}) and adding terms one finds that the second-order contribution to the energy density is proportional to 
$(U_{\rn\rp}^{S=0})^2+3(U_{\rn\rp}^{S=1})^2$.
The final result is
\begin{align}
&E^{(2)}= U_{\rn\rn}^2 \Phi(p_\rn, p_\rn)+ U_{\rp\rp}^2 \Phi(p_\rp, p_\rp)\nonumber\\
&+\left[ (U_{\rn\rp}^{S=0})^2 +3(U_{\rn\rp}^{S=1})^2\right]\Phi(p_\rn, p_\rp)\,,
\end{align}
where $\Phi$ is defined in Eq.~(\ref{Kanno}). Because this involves the same $\Phi$ function, the same discussion about the performance of the quadratic approximation applies to the case of neutron--proton mixtures (see Fig.~\ref{fig:quadratic} and discussion thereof).


On extending the calculations of Sec. \ref{qpinteraction} to the case of neutrons and protons with two spin components each, we find for the Landau parameters for neutron--proton mixtures:
\begin{align}
&f_{\rn\rn}^{(2)\up \up}=(U_{\rn\rn}^{\uparrow\downarrow})^2\chi^{\rn\rn}+ \left((U_{\rm np}^{\up\up})^2+(U_{\rm np}^{\up\down})^2\right)\chi^{\rp\rp}\,,\\
&f_{\rn\rn}^{(2)\up \down}=(U_{\rn\rn}^{\uparrow\downarrow})^2 (\Xi^{\rn\rn}+\chi^{\rn\rn})+(\tilde U_{\rm np}^{\up\down})^2 \chi^{\rp\rp}\,,\\
&f_{\rp\rp}^{(2)\up \up}=(U_{\rp\rp}^{\uparrow\downarrow})^2\chi^{\rp\rp}(q)+ \left((U_{\rm np}^{\up\up})^2+(U_{\rm np}^{\up\down})^2\right)\chi^{\rn\rn}\,,\\
&f_{\rp\rp}^{(2)\up \down}=(U_{\rp\rp}^{\uparrow\downarrow})^2 (\Xi^{\rp\rp}+\chi^{\rp\rp})+(\tilde U_{\rm np}^{\up\down})^2 \chi^{\rn\rn}\,,\\
&f_{\rn\rp}^{(2)\up \up}= (U_{\rn\rp}^{\up\up})^2\Xi^{\rn\rp} + \left((U_{\rn\rp}^{\up\up})^2+(\tilde U_{\rn\rp}^{\up\down})^2\right)\chi^{\rn\rp}\,,\\
&f_{\rn\rp}^{(2)\up \down}=\left((U_{\rn\rp}^{\uparrow\downarrow})^2 +(\tilde U_{\rn\rp}^{\up\down})^2\right)\Xi^{\rn\rp}+( U_{\rn\rp}^{\up\down})^2 \chi^{\rn\rp}\,.
\end{align}
The spin-symmetric and spin-antisymmetric interactions are given by $f_{\alpha\beta}^{\rs}=(f_{\alpha\beta}^{\up\up}+f_{\alpha\beta}^{\up\down})/2$ and $f_{\alpha\beta}^{\ra}=(f_{\alpha\beta}^{\up\up}-f_{\alpha\beta}^{\up\down})/2$ and therefore
\begin{align}
f_{\rn\rn}^{(2)\rm s} &=(U_{\rn\rn}^{\uparrow\downarrow})^2\left(\frac12 \Xi^{\rn\rn}+\chi^{\rn\rn}\right) \nonumber\label{fnns}\\
&+\left( \frac34(U_{\rm np}^{S=1})^2+\frac14 (U_{\rm np}^{S=0})^2 \right)\chi^{\rp\rp}\,,\\
f_{\rn\rn}^{(2)\rm a} &=-\frac12 (U_{\rn\rn}^{\uparrow\downarrow})^2 \Xi^{\rn\rn}
+\frac12 U_{\rm np}^{S=0}U_{\rm np}^{S=1} \chi^{\rp\rp} \,,\\
f_{\rn\rp}^{(2)\rm s} &=\left(\frac14 (U_{\rn\rp}^{S=0})^2+\frac34 (U_{\rn\rp}^{S=1})^2\right)(\Xi^{\rn\rp} +\chi^{\rn\rp})\,, \\
f_{\rn\rp}^{(2)\rm a}&=\frac14\left( (U_{\rn\rp}^{S=1})^2 -(U_{\rn\rp}^{S=0})^2 \right)\Xi^{\rn\rp} \nonumber \\
&+\frac12 U_{\rm np}^{S=1}\left(U_{\rm np}^{S=1} -U_{\rm np}^{S=0}\right) \chi^{\rn\rp}\,.\label{fnpa}
\end{align}
Landau parameters can be calculated from Eqs.~(\ref{fnns})--(\ref{fnpa}) by expanding the $\chi$'s and $\Xi$'s in terms of Legendre polynomials, see also the Appendix.

\section{Discussion and conclusions}

Low-density gases are model systems which have played an important role in studies of many-body physics. 
An interesting result of our work it that for low proton concentrations, the energy density has a term proportional to $n_\rp^{7/3} \ln n_\rp$, which is negative. For small $y=p_\rp/p_\rn$ one finds
\begin{multline}
\Phi \simeq \frac{mp_\rn^7}{3360\pi^6} \left[35 y^3 - 14 y^5 + \frac{70}{3} y^6 \right. \\
\left. - \frac{4}{105} (71 + 105 \ln[1/y]) y^7
- \frac72 y^8 + \frac23 y^9 + {\cal{O}}(y^{10})\right]. 
\label{Lowprotonconc}
\end{multline}
The $n_\rp^{7/3} \ln n_\rp$ term is not present in phenomenological energy density functionals such as those of the Skyrme type, where a term in the energy density proportional to a power greater than two in the total nucleon density $n$ is introduced in order to make the saturation density of nuclear matter finite. Typically this term varies as $n^{2+\alpha}$, with $\alpha = 1$ or $\alpha$ around $1/3$. This leads to the term beyond the quadratic one in an expansion of the energy density in powers of $n_\rp$ varying as $n_\rp^3$.

In recent work studying proton drip, it was convenient to use as the two quantities specifying the state of the system the proton density and the neutron chemical potential, rather than the nucleon densities \cite{Kelleretal}. It is straightforward to show that the $n_\rp^{7/3} \ln n_\rp$ terms in expansions of $E(n_\rn,n_\rp)$ and of $E(\mu_\rn,n_\rp)$ for small $n_\rp$ are identical.

For small differences between the neutron and proton densities, there is a term in the energy proportional to $(p_\rn-p_\rp)^4 \ln |p_\rn-p_\rp |$ (see, e.g., Ref.~\cite{Kaiseretal}). The corresponding term in the energy of a single-component spin-$1/2$ Fermi liquid as a function of the magnetic field $H$ leads to an $H^2\ln H$ term in the magnetic susceptibility which was invoked to explain measurements on palladium~\cite{Misawa}. 

For neutrons and protons, the low-density expressions for the energy density and other quantities are applicable only over a limited density range because the 
magnitudes of the scattering lengths are so large, $\sim 10$ fm, so the low-density expansion holds only for $\sim 10^{-3}$ fm$^{-3}$. An additional complication is that the neutron--proton scattering length in the triplet state is positive, which reflects the existence of a bound state, the deuteron. To take the deuteron and other bound states into account, one could extend the present calculation by including them explicitly, as in the virial expansion approach \cite{HorowitzSchwenk}. However, at densities greater than the typical density in the deuteron, one does not expect the bound states to play a dominant role and the low-density results we have derived should be a good approximation provided the scattering length in vacuo is replaced by an appropriate $S$-wave effective scattering length, as we remarked earlier. 

For a low-density mixture of neutrons and protons with a low proton concentration one finds that the energy density is given by Eq.~(\ref{Lowprotonconc}), which shows that the expansion parameter is $~(n_\rp/n_\rn)^{1/3}$. This is to be contrasted with the higher power that occurs with effective interactions of the Skyrme type.

In Ref.~\cite{Martikainen_et_al}, the authors, building on Refs.~\cite{GorkovMB, HeiselbergPSV}, calculated the influence of induced interactions on superfluid gaps in mixtures of fermions with three components, and this work could be extended to the case of four components using the results for induced interactions given above.

\acknowledgments

We are grateful to Emil Bjerrum-Bohr and Cristian Vergu for illuminating discussions on scattering amplitudes, to Piotr Chankowski and Pierbiagio Pieri for helpful correspondence regarding the work of Kanno~\cite{Kanno}, to Shin Takagi for facilitating contact to Kanno, to Silas Beane and Roland Farrell for discussions on the effective mass and pointing out the typo in Ref.~\cite{Platter2003}, and to Kai Hebeler for discussion on neutron--proton mixtures. Nordita is partially supported by Nordforsk. This work was supported in part by the European Research Council (ERC) under the European Union’s Horizon 2020 research and innovation programme (Grant Agreement No.~101020842).

\appendix

\section{Evaluation of integrals}
\label{details}

\subsection{Quasiparticle interaction}

Here we give some details of the calculations of the Landau quasiparticle interaction and the Landau parameters.

Equation (\ref{fabbasic}) for $f^{(2)\ra\rb}$ may be written in the form
\begin{align}
& f^{(2)\ra\rb}({\bf{p, p'}}) = f^{(2)\rb\ra}({\bf{p, p'}}) =U_0^2 (\Xi^{\ra\rb}+ \chi^{\ra\rb}) \,,
\label{f2ab}
\end{align}
where $\Xi^{\ra\rb}=\Xi^\ra + \Xi^\rb$ and $\chi^{\ra\rb}=\chi^\ra+\chi^\rb $. 
Here
\begin{align}
&\Xi^\ra(q )= -2mU_0^2\sum_{1}\frac{n_{1\ra}}{p_\ra^2 +p_\rb^2 -p_1^2 -(\kk-\pp_1)^2}\nonumber\\
&=\frac{m}{2\pi^2}\left[\frac{p_\ra}2-\frac{q}{8}\ln \left| \frac{(q+p_\ra)^2-p_\rb^2}{(q- p_\ra)^2-p_\rb^2} \right| \right. \nonumber \\
& \left. +\frac{\left(p_\ra^2 - p_\rb^2 \right)}{8k}
\ln \left| \frac{(k+p_\ra)^2-p_\rb^2}{(k-p_\ra)^2-p_\rb^2} \right| \right].
\label{xia}
\end{align}
The result for $\Xi^\rb$ is obtained from Eq.~(\ref{xia}) by interchanging a and b.
Thus one finds
\begin{align}
&\Xi^{\ra\rb}(q)
= \frac{m}{2\pi^2}\left[\frac{p_\ra +p_\rb}2 -\frac{q}{4}\ln \left| \frac{q+p_\ra+p_\rb}{q-p_\ra-p_\rb} \right| \right. \nonumber \\
& \left. + \frac{\left(p_\ra^2 - p_\rb^2 \right)}{4k}
\ln \left| \frac{k+p_\ra -p_\rb}{k-p_\ra +p_\rb} \right| \right].
\label{xiab}
\end{align}

The quantities $\chi^\ra$ and $\chi^\rb$ are given by
\begin{align}
&\chi^\ra(q)=\frac{m}{2\pi^2}
\left[ \frac{p_\ra}4 \left( 1+\frac{p_\ra^2-p_\rb^2}{q^2} \right)\right. \nonumber \\
&- \left. \frac{[(p_\ra^2-p_\rb^2)^2-2( p_\ra^2+p_\rb^2) q^2 +q^4]}{16 q^3} \ln \left| \frac{(p_\ra + q)^2-p_\rb^2}{(p_\ra-q)^2-p_\rb^2} \right| \right]
\end{align}
and the corresponding result with a and b interchanged.
Therefore
\begin{align}
&\chi^{\ra\rb}(q)
=\frac{m}{2\pi^2}
\left[ \frac{p_\ra+p_\rb}4 \left( 1+\frac{(p_\ra-p_\rb)^2}{q^2} \right)\right. \nonumber \\
&- \left. \left\{\frac{(p_\ra^2-p_\rb^2)^2}{8q^3}-\frac{( p_\ra^2+p_\rb^2)}{4q} +\frac{q}{8} \right\} \ln \left| \frac{q+p_\ra+p_\rb}{q-p_\ra-p_\rb} \right| \right]. 
\label{chiab}
\end{align}
For $p_\ra=p_\rb$, Eq.~(\ref{chiab}) reduces to the static Lindhard function Eq.~(\ref{staticLindhard}).
Insertion of Eqs.~(\ref{xiab}) and (\ref{chiab}) into Eq.~(\ref{f2ab}) leads to Eq.~(\ref{fabq}).

\subsection{Landau parameters}

From Eq.~(\ref{LandauParam}) it follows that Landau parameters are related to the quasiparticle interaction by the expression
\be
\frac{f_\ell^{\alpha\beta}}{2\ell+1}=\frac12 \int_{-1}^1 d\zeta\, f^{\alpha\beta}(q)P_\ell(\zeta)\,.
\ee
The quantity $\zeta=\hat\pp\cdot\hat\pp'$ is related to either $q$ or $k$ by the relations
\be
\zeta=\frac{p_\alpha^2+p_\beta^2-q^2}{2p_\alpha p_\beta}=\frac{k^2-p_\alpha^2-p_\beta^2}{2p_\alpha p_\beta}\,,
\ee
which follow from Eqs.~(\ref{q}) and (\ref{k}) for $\pp$ and $\pp'$ lying on the Fermi surface for the species in question.
Thus one finds
\be
f_0^{(2)\ra\ra}=\frac{m U_0^2}{2\pi^2} \int_0^{2p_\ra} \frac{dq q}{2p_\ra^2}\left[ \frac{ p_\rb}2+\left(\frac{p_\rb^2}{2q}-\frac{q}{8} \right)\ln\left| \frac{q+2p_\rb}{q-2p_\rb} \right| \right]
\label{faa0}
\ee
and
\begin{align}
\frac{1}{3}f_1^{(2)\ra\ra}=\frac{m U_0^2}{2\pi^2} \int_0^{2p_\ra} \frac{dq }{2p_\ra^2} &\left[ \left(\frac{p_\rb^2}{2}-\frac{q^2}{8} \right) \right. \nonumber \\
&\times \left.\left( 1-\frac{q^2}{2p_\ra^2}\right) \ln\left| \frac{q+2p_\rb}{q-2p_\rb} \right| \right].
\label{faa1}
\end{align}
On performing the integrals, one arrives at Eqs.~(\ref{faa0final}) and (\ref{faa1final}).
Results for the b--b interaction are obtained from those for the a--a interaction by interchanging a and b in Eqs.~(\ref{faa0}) and~(\ref{faa1}).

For the Landau parameters for particles in different internal states, it follows from Eq.\ (\ref{fabq}) that
\begin{align}
&f^{(2)\ra\rb}_0=\frac{mU_0^2}{2\pi^2}\int_{-1}^1\frac{d \zeta}{2}
\left[ \frac{p_\ra+p_\rb}4 \left( 3+\frac{(p_\ra-p_\rb)^2}{q^2} \right)\right. \nonumber \\
&- \left(\frac{(p_\ra^2-p_\rb^2)^2}{8q^3}-\frac{( p_\ra^2+p_\rb^2)}{4q} +\frac{3 q}{8} \right) \ln \left| \frac{q+p_\ra+p_\rb}{q-p_\ra-p_\rb} \right| \nonumber \\
&+ \left.\frac{\left(p_\ra^2 - p_\rb^2 \right)}{4k}
\ln \left| \frac{k+p_\ra -p_\rb}{k-p_\ra +p_\rb} \right| \right] .
\end{align}
Consequently,
\begin{align}
&f^{(2)\ra\rb}_0=\frac{mU_0^2}{2\pi^2}
\left[ \frac{3}{4}(p_\ra+p_\rb) \right.\nonumber \\
&+ \frac{(p_\ra +p_\rb)(p_\ra-p_\rb)^2}{8p_\ra p_\rb} \ln\left|\frac{p_\ra + p_\rb}{p_\ra -p_\rb} \right| 
- \frac{(p_\ra^2-p_\rb^2)^2}{16p_\ra p_\rb} I_{-2} \nonumber \\ & \left.+ \frac{( p_\ra^2+p_\rb^2)}{8 p_\ra p_\rb}I_0 -\frac{3 }{16 p_\ra p_\rb} I_2 
+ \frac{\left(p_\ra^2 + p_\rb^2 \right)}{8 p_\ra p_\rb}\tilde I_0 \right],
\label{fab0}
\end{align}
where
\be
I_n =\int_{|p_\ra -p_\rb|}^{p_\ra + p_\rb} d y \,y^n \ln \left| \frac{y+p_\ra +p_\rb}{y-p_\ra -p_\rb} \right|
\ee
and
\be
\tilde I_n =\int_{|p_\ra -p_\rb|}^{p_\ra + p_\rb} d y\, y^n \ln \left| \frac{y+p_\ra -p_\rb}{y-p_\ra +p_\rb} \right|.
\ee
Also
\begin{align}
&\frac{f^{(2)\ra\rb}_1}{3}=\frac{mU_0^2}{2\pi^2}
\left[ -\frac{(p_\ra +p_\rb) (p_\ra- p_\rb)^2}{8 p_\ra p_\rb}\right.\nonumber \\
&+ \frac{(p_\ra +p_\rb)(p_\ra-p_\rb)^2(p_\ra^2 + p_\rb^2)}{16p_\ra^2 p_\rb^2} \ln\left|\frac{p_\ra + p_\rb}{p_\ra -p_\rb} \right| \nonumber \\
& - \frac{(p_\ra^2-p_\rb^2)^2(p_\ra^2 + p_\rb^2)}{32p_\ra^2 p_\rb^2} I_{-2} 
+ \frac{( 3p_\ra^4+2p_\ra^2 p_\rb^2+3p_\rb^4)}{32 p_\ra^2 p_\rb^2}I_0 \nonumber \\ 
& -\frac{5 }{32}\frac{(p_\ra^2 + p_\rb^2)}{ p_\ra^2 p_\rb^2} I_2 +\frac{3}{32p_\ra^2 p_\rb^2}I_4
- \frac{(p_\ra^4 - p_\rb^4 )}{16 p_\ra^2 p_\rb^2}\tilde I_0 \nonumber \\
&\left. + \frac{(p_\ra^2-p_\rb^2)}{16 p_\ra^2 p_\rb^2} \tilde I_2 \right] .
\label{fab1}
\end{align}
Specifically,
\begin{align}
&I_{-2}= \frac{2}{p_\ra^2 -p_\rb^2}\left[ -p_\ra\ln\left|\frac{p_\ra- p_\rb}{p_\ra}\right|+ p_\rb\ln\left|\frac{p_\ra- p_\rb}{p_\rb}\right| \right] , \\
&I_0= 2\left[p_\ra\ln\left(\frac{p_\ra + p_\rb}{p_\ra}\right)+ p_\rb\ln\left(\frac{p_\ra + p_\rb}{p_\rb}\right) \right] , \\
& I_2= \frac{2}{3}\left[p_\ra(p_\ra^2+3 p_\rb^2)\ln\left(\frac{p_\ra + p_\rb}{p_\ra}\right) \right. \nonumber\\
&+ \left. p_\rb(p_\rb^2+3 p_\ra^2)\ln\left(\frac{p_\ra + p_\rb}{p_\rb}\right) +2p_\ra p_\rb(p_\ra +p_\rb) \right] , \\
&I_4= \frac{2}{5}\left[ p_\ra(p_\ra^4+10p_\ra^2 p_\rb^2+5p_\rb^4)\ln\left(\frac{p_\ra + p_\rb}{p_\ra}\right) \right. \nonumber\\
&+ p_\rb(p_\rb^4+10p_\ra^2 p_\rb^2+5p_\ra^4)\ln\left(\frac{p_\ra + p_\rb}{p_\rb}\right) \nonumber \\
&\left. + 4p_\ra p_\rb(p_\ra +p_\rb)(p_\ra^2 + p_\ra p_\rb + p_\rb^2) \right] ,\\
&\tilde I_0= 2\left[-p_\ra\ln\left|\frac{p_\ra - p_\rb}{p_\ra}\right|+ p_\rb\ln\left|\frac{p_\ra - p_\rb}{p_\rb}\right| \right] , \\
&\tilde I_2= \frac{2}{3}\left[-p_\ra(p_\ra^2+3 p_\rb^2)\ln\left|\frac{p_\ra - p_\rb}{p_\ra}\right| \right. \nonumber\\
&+ \left. p_\rb(p_\rb^2+3 p_\ra^2)\ln\left|\frac{p_\ra -p_\rb}{p_\rb}\right| +2p_\ra p_\rb(p_\ra -p_\rb) \right] .
\end{align}
Simplifying expressions (\ref{fab0}) and (\ref{fab1}) gives Eqs.~(\ref{fab0final}) and~(\ref{fab1final}).

For the case of equal Fermi momenta, $p_\ra=p_\rb=p_{\rm F}$, one finds $f^{(2)\ra\rb}_0=2(mp_{\rm F}U_0^2/2\pi^2)$ and $\frac13{f^{(2)\ra\rb}_1}=(mp_{\rm F}U_0^2/2\pi^2) (1+8\ln 2)/15$, in agreement with what is obtained by taking the angular moments of the expressions in Ref.~\cite[Eq.~(6.16)]{LifshitzPitaevskii}.


\begin{thebibliography}
\bibliographystyle{}
\bibitem{StoneReinhard} For a review of Skyrme energy functionals, see J. R. Stone and P.-G. Reinhard, The Skyrme interaction in finite nuclei and nuclear matter, Prog. Part. Nucl. Phys. {\bf 58}, 587 (2007).
\bibitem{Stoner} E. C. Stoner, LXXX. Atomic moments in ferromagnetic metals and alloys with non-ferromagnetic elements, Phil. Mag. {\bf 15}, 1018 (1933).
\bibitem{Landau1} L. D. Landau, The theory of a Fermi liquid, Sov. Phys. JETP {\bf 3}, 920 (1957).
\bibitem{Landau2} L. D. Landau, Oscillations in a Fermi liquid, Sov. Phys. JETP {\bf 5}, 101 (1957).
\bibitem{Landau3} L. D. Landau, On the theory of the Fermi liquid Sov. Phys. JETP {\bf 8}, 70 (1959).
\bibitem{BaymPethick} G. Baym and C. J. Pethick, Landau Fermi Liquid Theory: Concepts and Applications (Wiley, New York, 1991).
\bibitem{Migdal} A. B. Migdal, Theory of Finite Fermi Systems and Applications to Atomic Nuclei (Interscience, New York, 1967).
\bibitem{FrimanHebelerSchwenk} B. Friman, K. Hebeler, and A. Schwenk, Renormalization group and Fermi liquid theory for many-nucleon systems, Lect. Notes Phys. {\bf 852}, 245 (2012).
\bibitem{Kanno} S. Kanno, Criterion for the ferromagnetism of hard sphere Fermi liquid, Prog. Theor. Phys. {\bf 44}, 813 (1970).
\bibitem{HuangYang} K. Huang and C. N. Yang, Quantum-mechanical many-body problem with hard-sphere interaction, Phys. Rev. {\bf 105}, 767 (1957).
\bibitem{DeDominicisMartin} C. De Dominicis and P. C. Martin, Energy of interacting Fermi systems, Phys. Rev. {\bf 105}, 1417 (1957). 
\bibitem{AbrikosovKhalatnikov} A. A. Abrikosov and I. M. Khalatnikov, Concerning a model for a non-ideal Fermi gas, Sov. Phys. JETP {\bf 6}, 888 (1958). 
\bibitem{LifshitzPitaevskii} E. M. Lifshitz and L. P. Pitaevskii, \textit{Statistical Physics}, Part 2, Second edition, (Pergamon, Oxford, 1980), \S 6.
\bibitem{HammerFurnstahl} H.-W. Hammer and R. J. Furnstahl, Effective field theory for dilute Fermi systems, Nucl. Phys. A {\bf 678}, 277 (2000).
\bibitem{WDS2020} C. Wellenhofer, C. Drischler, and A. Schwenk, Dilute Fermi gas at fourth order in effective field theory, Phys. Lett. B \textbf{802}, 135247 (2020).
\bibitem{WDS2021} C. Wellenhofer, C. Drischler, and A. Schwenk, Effective field theory for dilute Fermi systems at fourth order, Phys. Rev. C \textbf{104}, 014003 (2021).
\bibitem{FratiniPilati} E. Fratini and S. Pilati, Zero-temperature equation of state and phase diagram of repulsive fermionic mixtures, Phys. Rev. A {\bf 90}, 023605 (2014).
\bibitem{ChankowskiWojtkiewicz1} P. Chankowski and J. Wojtkiewicz, 
Ground-state of the polarized dilute gas of interacting 
spin-$\frac12$ fermions,
Phys. Rev. B {\bf 104}, 144425 (2021).
\bibitem{ChankowskiWojtkiewicz2} P. Chankowski and J. Wojtkiewicz, On the ground-state energy of a mixture of two different oppositely polarized fermionic gases, Acta Phys. Pol. {\bf 53}, 1 (2022).
\bibitem{Pera} J. Pera, J. Casulleras, and J. Boronat, Itinerant ferromagnetism in dilute SU(N) Fermi gases, SciPost Phys. {\bf 14}, 038 (2023).
\bibitem{Zwerger} W. Zwerger (Ed.), The BCS-BEC Crossover and the Unitary Fermi Gas (Springer, Berlin, Heidelberg, 2012).
\bibitem{Sjoeberg} O. Sj\"oberg, On the Landau effective mass in asymmetric nuclear matter, Nucl. Phys. A {\bf 265}, 511 (1976).
\bibitem{Galitskii} V. M. Galitskii, The energy spectrum of a non-ideal Fermi gas, Sov. Phys. JETP 34, 104 (1958).
\bibitem{Platter2003} L. Platter, H.~W. Hammer, and U. G. Mei{\ss}ner, Quasiparticle properties in effective field theory, Nucl. Phys. A \textbf{714}, 250 (2003).
\bibitem{Kelleretal} J. Keller, K. Hebeler, C. J. Pethick, and A. Schwenk, Neutron star matter as a dilute solution of protons in neutrons, Phys. Rev. Lett. {\bf 132}, 232701 (2024).
\bibitem{Kaiseretal} C. Wellenhofer, J. W. Holt, and N. Kaiser, Divergence of the isospin-asymmetry expansion of the nuclear equation of state in many-body perturbation theory, Phys. Rev. C \textbf{93}, 055802 (2016).
\bibitem{Misawa} S. Misawa, Logarithmic field dependence of the susceptibility of a paramagnetic Fermi liquid -- the Pd problem, Phys. Rev. Lett. {\bf 26}, 1632 (1971).
\bibitem{HorowitzSchwenk} C. J. Horowitz and A. Schwenk, Cluster formation and the virial equation of state of low-density nuclear matter, Nucl. Phys. A {\bf 776}, 55 (2006).
\bibitem{Martikainen_et_al} J.-P. Martikainen, J. J. Kikkunen, P. T\"orma, and C. J. Pethick, Induced interactions and the superfluid transition temperature
in a three-component Fermi gas, Phys. Rev. Lett. {\bf 103}, 260403 (2009).
\bibitem{GorkovMB} L. P. Gor'kov and T. K. Melik-Barkhudarov, Contribution to the theory of superfluidity in an imperfect Fermi gas, Sov. Phys. JETP {\bf 13}, 1018 (1961).
\bibitem{HeiselbergPSV} H. Heiselberg, C. J. Pethick, H. Smith, and L. Viverit, Influence of induced interactions on the superfluid transition in dilute Fermi gases. Phys. Rev. Lett. {\bf 85}, 2418 (2000).

\end{thebibliography}
\end{document}